# Fundamental inequalities in the Stoner-Wohlfarth model

C. A. Iglesias[*], J. C. R. de Araújo, E. F. Silva, M. Gamino, M. A. Correa, and F. Bohn[†]

*Departamento de Física, Universidade Federal do Rio Grande do Norte, 59078-900 Natal, Rio Grande do Norte, Brazil*

We report two fundamental inequalities in the Stoner-Wohlfarth model. Specifically, we investigate the theoretical limit for the initial magnetic susceptibility in a system described by the Stoner-Wohlfarth approach. We also find analytical solutions for the magnetization in the low-field regime and obtain the borderline value for the uniaxial anisotropy constant in such an ideal Stoner-Wohlfarth system. We go beyond and introduce a general mean-field theory for interacting Stoner-Wohlfarth-like systems, thus estimating how the initial magnetic susceptibility is affected by the dipolar and exchange interactions inside the system. By means of a simple insight from a fundamental inequality for the magnetic susceptibility of an ideal Stoner-Wohlfarth system, we show its violation is a signature of the existence of exchange interactions between nanoparticles in an interacting Stoner-Wohlfarth-like system.

## I. INTRODUCTION

For more than a century, the dependence of magnetization on magnetic field has drawn attention as a signature of magnetic materials, bringing fundamental insights into the physical mechanisms involved in magnetization dynamics. On the theoretical side, approaches have been developed to address the magnetic properties of magnets, capturing essential features of the magnetization process [1–7]. One celebrated example that has shaped our thinking despite its simplicity is the Stoner-Wohlfarth (SW) model [8]. This model represents a theoretical approach often used to simulate the expected magnetic properties of noninteracting uniaxial anisotropy blocked particle systems. The random *SW model* appears to be a very useful tool to predict important parameters observed in real *noninteracting systems*. For *interacting systems* in turn, the presence of dipolar and/or exchange interactions affecting the magnetization dynamics makes the description of all the magnetic properties a hard task.

Within this framework, magnetic nanoparticles and their wide diversity provide a fascinating playground for theoretical and experimental investigations of the magnetic properties in both noninteracting and interacting systems. From a historical point of view, magnetic nanoparticles have been the focus of numerous studies for several decades due to their challenging physical properties and potential for application [9,10]. Perhaps most of the progress on the discovery of novel materials and the exploration of the dynamic magnetic response in diverse particle systems has been driven by the technological demand. Within this perspective, magnetic nanoparticles have appeared, for instance, in the context of biomedical engineering [11–18], as well as in a wide variety of technological applications [19–22]. Nevertheless, recent advances in the field of magnetization dynamics have stimulated renewed interest in phenomena involving the interactions between magnetic nanoparticles. For magnetic systems, it is well known that their magnetic properties are dependent on numerous issues, including experimental parameters employed in the production of the sample and features owing to the chemical composition [14,23–30]. For nanoparticles, in addition, the interactions between particles have a key role in all the magnetic properties and magnetization dynamics [4,31–39]. The presence and intensity of such interactions have often been explored by employing remanence plots [40–44]. While several aspects of the interactions have been the subject of recent analysis, it remains unclear whether or not there is a simple, straightforward way to relate the general features of a system, such as the magnetic susceptibility, and the type of interaction between particles contributing to its properties and dynamics.

In this paper, we derive two fundamental inequalities in the context of the Stoner-Wohlfarth model. We investigate the theoretical limit for the initial magnetic susceptibility in a system described by the SW approach, find analytical solutions for the magnetization in the low-field regime, and obtain the borderline value for the uniaxial anisotropy constant in such an ideal SW system. Going further, we introduce a general mean-field theory for interacting SW-like systems, and we also estimate how the initial magnetic susceptibility is affected by the dipolar and exchange interactions inside the system. Then, using a fundamental inequality for the magnetic susceptibility of an ideal SW system, we show its violation is a signature of the existence of exchange interactions between nanoparticles in an interacting Stoner-Wohlfarth-like system.

## II. THEORETICAL APPROACH AND DISCUSSION

In this section, we initially give a brief overview of the Stoner-Wohlfarth model to establish the basis of our model and make clear the nomenclature employed in this work. Here, it is worth mentioning that the term *Stoner-Wohlfarth-like system* (SW-like system) is widely taken in the construction of our ideas and refers to a system consisting of interacting (or not) uniaxial anisotropy single-domain blocked particles. Moving forward, we build some well-known work-energy relations involved in the magnetization processes considered in our context. From this framework, we first provide a fundamental inequality for the initial magnetic susceptibility that is valid for ideal SW systems; in the sequence we find specific analytical solutions for the magnetization at low fields in the SW model, then provide a second fundamental inequality, and introduce a mean-field theory for interacting SW systems. Finally, we suggest the violation of the fundamental inequality for the magnetic susceptibility is a signature of the major contribution of ferromagnetic exchange interactions between blocked nanoparticles in a SW-like system.

---

[*]iglesias@fisica.ufrn.br
[†]felipebohn@fisica.ufrn.br

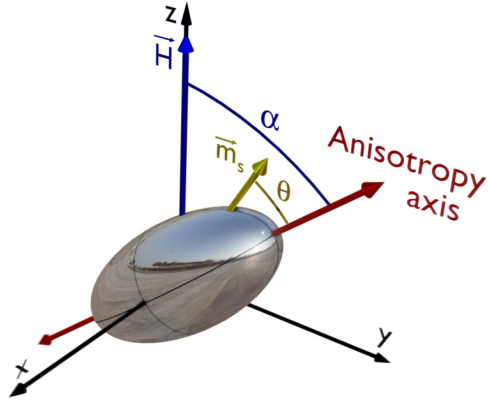

FIG. 1. Schematic diagram of the theoretical system. Here, we show just a single particle with uniaxial magnetic anisotropy (double arrow). We consider $\vec{H}$ to be the magnetic field vector and $\vec{m}_s$ to be the magnetization vector, the latter having an amplitude equal to the saturation magnetization. Thus, $\alpha$ is the angle between $\vec{H}$ with respect to the uniaxial magnetic anisotropy, while $\theta$ is the angle between $\vec{m}_s$ and the direction of the anisotropy.

It is worth pointing out that we use, in the construction of the theoretical approach, conventional SI units for all quantities; they are summarized in Table I in the Appendix.

### A. A brief recapitulation of the SW model

We start our approach by recalling the SW model. It assumes that the system is composed of a set of noninteracting, uniaxial anisotropy single-domain particles at a temperature of 0 K; that is, the particles are completely blocked magnetically. The reversal of the magnetization is due to coherent rotation of the single-domain magnetic particles, and thermal effects on magnetization are neglected [8].

Figure 1 shows a sketch of the theoretical system using the SW model, together with the definitions of the relevant vectors and angles considered in our approach. In this case, we assume $\vec{H}$ is the magnetic field vector and $\vec{m}_s$ is the magnetization vector whose amplitude is the saturation magnetization; additionally, we consider $\alpha$ to be the angle of $\vec{H}$ with respect to the uniaxial magnetic anisotropy, and $\theta$ corresponds to the angle between $\vec{m}_s$ and the direction of the anisotropy.

The SW model is based on the minimization of the free energy of the system. From the appropriate magnetic free energy, a routine for minimization determines the values of the equilibrium angle $\theta_m$ of the saturation magnetization vector for a given magnetic field; and from this procedure we may obtain the magnetization curve, i.e., the set of values of the component of the magnetization along $\vec{H}$ for each field value.

For the sake of simplicity, the SW model takes into account only the Zeeman interaction and the effective uniaxial magnetic anisotropy terms. The Zeeman energy per particle is

$$E_Z = -\mu_0 m_s V_p H \cos(\alpha - \theta), \quad (1)$$

while the energy associated with the uniaxial magnetic anisotropy of a particle is

$$E_a = k_{\text{eff}} V_p \sin^2(\theta), \quad (2)$$

where $\mu_0$ is the magnetic permeability of the free space, $H$ is the amplitude of the magnetic field, $m_s$ is the volumetric saturation magnetization, $V_p$ is the volume of the particle, $k_{\text{eff}}$ is the effective uniaxial magnetic anisotropy constant, and $\alpha$ and $\theta$ are the aforementioned angles. The magnetic free energy $E_f$ per particle in this case is given by

$$E_f = k_{\text{eff}} V_p \sin^2(\theta) - \mu_0 m_s V_p H \cos(\alpha - \theta). \quad (3)$$

Notice that the magnetic field $\vec{H}$, applied along a given direction described by the angle $\alpha$ with respect to the anisotropy axis, rotates the magnetization vector $\vec{m}_s$ at an angle $\theta$ from the orientation of the anisotropy. It gives rise to a restoring force due to the anisotropy against the magnetization rotation. Hence, the equilibrium state of the magnetization, represented by the equilibrium angle $\theta_m$, is obtained by minimizing the magnetic free energy for each $H$ value [45], i.e.,

$$\partial E_f / \partial \theta = 0, \quad (4)$$

whose solution $\theta_m$ must satisfy the condition

$$\partial^2 E_f / \partial \theta^2 > 0. \quad (5)$$

Using Eqs. (3) and (4), we can find the following relation:

$$2 k_{\text{eff}} \sin(\theta) \cos(\theta) = \mu_0 m_s H \sin(\alpha - \theta). \quad (6)$$

Additionally, we can express the volumetric magnetization $m$, i.e., the component of $\vec{m}_s$ along the $\vec{H}$, as

$$m = m_s \cos(\alpha - \theta). \quad (7)$$

Here, we raise some well-known predictions for the SW model that are of interest. First, for now, let us consider the simplest case in which the field is perpendicular to the anisotropy axis, a situation represented by $\alpha = \pi/2$. By combining Eqs. (6) and (7) and using the identity $\cos(\pi/2 - \theta) = \sin(\theta)$, we obtain

$$\frac{m}{m_s} = \frac{\mu_0 m_s}{2 k_{\text{eff}}} H, \quad (8)$$

which reveals the magnetization has a linear dependence on the field for $\alpha = \pi/2$. In this condition, the magnetization saturates ($m/m_s = 1$) when the magnetic field reaches a value of

$$H_k = \frac{2 k_{\text{eff}}}{\mu_0 m_s}, \quad (9)$$

which is known as the anisotropy field $H_k$.

Second, let us now consider a system composed of a set of randomly oriented particles with uniaxial magnetic anisotropy. In this case, the net value of the volumetric magnetization $m$ is obtained through the average value [8]

$$\langle m(H) \rangle = m_s \frac{\int_0^{\pi/2} \cos[\alpha - \theta_m(\alpha)] f(\alpha) \, 2\pi \, \sin(\alpha) \, d\alpha}{\int_0^{\pi/2} f(\alpha) \, 2\pi \, \sin(\alpha) \, d\alpha}, \quad (10)$$

where $f(\alpha)$ is a distribution function that describes the angular distribution of the magnetic anisotropy per particle. The limits of the integrals are 0 and $\pi/2$ rad due to the symmetry of the uniaxial anisotropy. For the random SW case, $f(\alpha) = 1$, and the anisotropy field is understood as the minimum field value needed to cause an irreversible magnetization reversal of all particles composing the system. Additionally, the model also provides normalized ratios for remanent magnetization $m_r$ and coercive field $H_c$. Specifically, we find $m_r/m_s = 0.5$ and $H_c/H_k = 0.48$ from the magnetization curve taken for the random SW system. Such well-known relations are very useful given they may be interpreted as pattern parameters to study magnetic samples that are, in principle, thought to be SW-like systems.

### B. Work-energy relations

Now, we bring in some well-known work-energy relations involved in the magnetization process [45]. The relations between energy and work are achieved from a macroscopic point of view and are of the utmost importance for evaluating the energy contributions from the interactions between nanoparticles in a collective system.

In order to estimate the work undergone by magnetic nanoparticles in a magnetization process, let us suppose a system (a set of noninteracting nanoparticles) consists of a magnetic sample with a cylindrical form, with radius $R$ and length $l$; moreover, let us consider a solenoid with $N$ turns, radius $R$, and length $l$, wound around the sample. For the sake of simplicity, we assume $l \gg R$, so that it may be taken as an ideal solenoid. Under these assumptions, the magnitude of the magnetic field $H$ produced inside the solenoid due to the electrical current $i$ flowing through the wire may be written as [46]

$$H = \frac{Ni}{l}. \quad (11)$$

The solenoid is connected to an electrical power source, which provides an electromotive force in a time interval $dt$ and produces an electrical current variation $di$ in the solenoid. In the magnetization process, as soon as the current is modified by an amount $di$, we observe a change in magnetic flux $d\Phi$ inside the solenoid and, consequently, throughout the sample. Keeping in mind we still maintain the ideal conditions, assuming the field is homogeneous through the cross-sectional area $A$ of the sample, the magnetic flux may be simply written as $\Phi = AB$, where $B$ is the magnetic induction. Then, the change in magnetic flux inside the solenoid may be expressed as

$$\frac{d\Phi}{dt} = A \frac{dB}{dt}. \quad (12)$$

Such a change in the flux causes a back electromotive force $\varepsilon_b$ in the solenoid, which by means of the Faraday's law [46] is

$$\varepsilon_b = -N d\Phi/dt, \quad (13)$$

and work must be done to overcome this back electromotive force [45].

In this framework, the work per unit volume $w$ done by the electrical power source in a time interval $dt$ against the back electromotive force on the solenoid wire can be written in differential form as

$$dw = \frac{1}{V_{\text{in}}} P dt = \frac{1}{V_{\text{in}}} Vi \, dt, \quad (14)$$

where $V_{\text{in}}$ is the inner volume of the solenoid and $P = Vi$ is the electrical power delivered by the source to the circuit. Notice that $i$ is the final electrical current value in the time interval $dt$, and $V = -\varepsilon_b$ is the voltage between the terminals of the solenoid. Since we are interested only in the magnetic effects, we neglect in Eq. (14) the electrical resistance of the solenoid and the corresponding Joule effect contribution to the energy balance.

From Eq. (14) and taking into account Eqs. (11), (12), and (13), the electrical work per unit volume done by the electrical source can be expressed in a generalized differential form as

$$dw = H \, dB, \quad (15)$$

with $B = \mu_0(H + m)$. Without loss of generality, here, we use simply the magnitudes $H$ and $B$, assuming the fact that the magnetization vector $\vec{m}$ is the volumetric average of the magnetic moment of the whole sample along the direction defined by the external magnetic field vector $\vec{H}$.

From this, we can explore straightforwardly some cases of interest. The first consists of the case in which the solenoid is empty, without the presence of the sample. Here, the total electrical work per unit volume done by the electrical power source is totally converted to potential energy, which remains stored in the magnetic field. Then, assuming $B = \mu_0 H$ and integrating Eq. (15) from $H = 0$ up to a given $H$ value, we may express the conservative energy per unit volume stored in the magnetic field $u_m$ as

$$u_m = \tfrac{1}{2} \mu_0 H^2. \quad (16)$$

Next, the second case is the one in which the sample is inside the solenoid. By taking into account the general principle of energy conservation, we can split the total electrical work per unit volume done by the electrical source into two components, the aforementioned energy per unit volume stored in the magnetic field $u_m$ and the work per unit volume undergone by the sample $w_s$. Hence, in differential form, $dw_s = dw - du_m$, which becomes, after integration,

$$w_s = \mu_0 \int H \, dm. \quad (17)$$

This equation is a general expression for obtaining the work undergone by a noninteracting magnetic nanoparticle system due to its interaction with the magnetic field.

### C. Fundamental inequalities

For our purposes, it becomes important to demonstrate that Eq. (16) describing the conservative energy per unit volume stored in the magnetic field represents precisely the maximum work per unit volume that can be done by the electrical power source of the experimental setup in a magnetization process.

As previously mentioned, the electrical resistance of the solenoid was neglected in the calculation of Eq. (14); hence, the electrical current in the solenoid may be written as [47]

$$i = \frac{V}{X_L} = \frac{V}{\omega L}, \quad (18)$$

where $X_L = \omega L$ is the inductive reactance, with $\omega$ being the angular frequency associated with the electric current variation and $L$ corresponding to the general inductance of the solenoid. Notice that here we are considering only the initial transient regime of the electrical current, where the electrical power source does work due to the emergence of inductive reactance in the solenoid.

The general inductance $L$ of the solenoid can be expressed in terms of $L_0$, i.e., the inductance when the solenoid is empty, as

$$L = \mu_r L_0 = \frac{\mu}{\mu_0} L_0, \quad (19)$$

where $\mu_r = \mu/\mu_0$ is the relative magnetic permeability, with $\mu_0$ being the magnetic permeability of the free space and $\mu$ corresponding to the magnetic permeability of the material inside the solenoid.

It is worth pointing out that magnetic materials have the inequality $\mu \geqslant \mu_0$. As a result, taking into account Eqs. (18) and (19), the work per unit volume done by the electrical power source in a time interval $dt$ against the back electromotive force on the solenoid wire given by Eq. (14) may be rewritten as

$$dw = \frac{1}{V_{\text{in}}} \frac{V^2}{\omega \mu_r L_0} dt. \quad (20)$$

Therefore, we conclude that the maximum $dw$ takes place when the condition $\mu = \mu_0$ is satisfied, and it corresponds specifically to the case in which the solenoid is empty. Within this context, we confirm that the conservative energy per unit volume stored in the magnetic field given by Eq. (16) indubitably does represent the maximum work per unit volume that can be done by the electric power source.

Once the work undergone by a macroscopic magnetic system and the maximum work that can be done in a magnetization process are established, we address the work undergone by an ideal system described by the SW model. For a system consisting of noninteracting, uniaxial anisotropy single-domain blocked particles, the work in the magnetization process in the low-field regime is done against only the individual restoring force arising in each particle due to the orientation of the magnetic field in a nonparallel direction with respect to the anisotropy axis. In the linear magnetization regime, in which $m = \chi H$, with $\chi$ being the initial magnetic susceptibility, such work, after integration of Eq. (17) from $m = 0$ to a given $m$ value, comes to

$$w_s = \tfrac{1}{2} \mu_0 m H. \quad (21)$$

Here, we must remark that the work undergone by a SW system represented by Eq. (21) is valid just for the conditions in which the magnetic field is well below both the coercive field $H_c$ and anisotropy field $H_k$, i.e., $H \ll H_c$ and $H \ll H_k$.

Next, from the definition of the initial magnetic susceptibility,

$$\chi = m/H, \quad (22)$$

and by means of the simple insight of multiplying the numerator and denominator in r.h.s. of Eq. (22) by the factor $(1/2)\mu_0 H$, we may express it as

$$\chi = \frac{\tfrac{1}{2}\mu_0 m H}{\tfrac{1}{2}\mu_0 H^2}. \quad (23)$$

Therefore, in the context of the SW model, we verify the initial magnetic susceptibility may be written as the ratio of the work undergone by the system [Eq. (21)] to the maximum work that can be done by the electrical power source in the magnetization process [Eq. (16)], i.e.,

$$\chi = \frac{w_s}{u_m}. \quad (24)$$

In this case, taking into account the general principle of energy conservation, we conclude that

$$\chi \leqslant 1 \quad (25)$$

for the magnetization curve in the linear regime. Equation (25) corresponds to the first fundamental inequality we aim to address here. It is, in principle, valid for any noninteracting single-domain (macrospin) blocked system and represents the theoretical limit of the initial magnetic susceptibility in an ideal nanoparticle system described by the SW model.

It is interesting to rewrite the maximum magnetic susceptibility in terms of the $m_s$ and $k_{\text{eff}}$ parameters. To this end, a possible way to investigate the initial magnetic susceptibility in uniaxial anisotropy systems is through the evaluation of the torques involved in the magnetization process. First, let us consider the low-field regime ($H \ll H_k$), in which we assume the approximation of $\theta \to 0$, i.e., $\vec{m}_s$ lies along the direction of the uniaxial magnetic anisotropy, as we can identify from Fig. 1. Under this condition, the torque exerted by the magnetic field on the macrospin, $\partial E_Z/\partial \theta$, is counterbalanced by the torque resulting from the uniaxial anisotropy, $\partial E_a/\partial \theta$ [48]. Then, from Eq. (2), we find

$$\frac{\partial E_a}{\partial \theta} = k_{\text{eff}} V_p \sin(2\theta) = -\frac{\partial E_Z}{\partial \theta}. \quad (26)$$

The initial torque of the magnetic field, taking into account the approximation $\theta \to 0$ and using Eq. (1), may be written as

$$\left(\frac{\partial E_Z}{\partial \theta}\right)_{\theta=0} = -\mu_0 m_s V_p H \sin(\alpha). \quad (27)$$

From Eqs. (26) and (27) and expressing $\sin(2\theta) = 2\theta$, which is valid for small $\theta$ values, we achieve

$$\theta = \frac{\mu_0 m_s H}{2 k_{\text{eff}}} \sin(\alpha), \quad (28)$$

and by means of Eq. (7), we obtain an analytical solution for the magnetization in the SW model considering the low-field

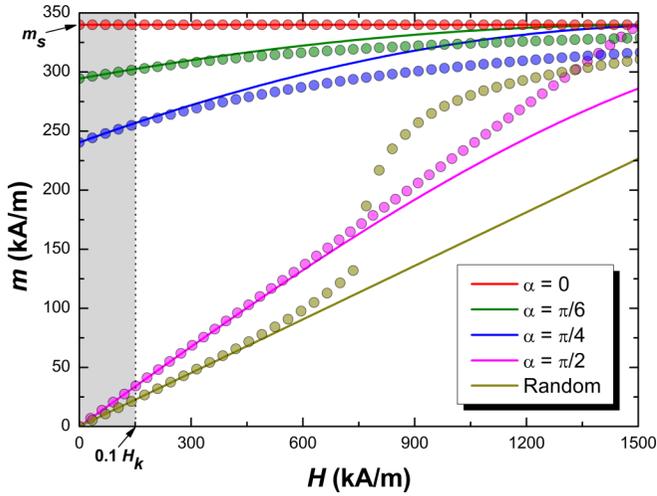

FIG. 2. Magnetization $m$ as a function of the magnetic field $H$, taking into account $\theta = 0$ as the initial situation, for selected $\alpha$ values. The solid lines are the analytical solutions given by Eq. (29) for the distinct $\alpha$ cases and by Eq. (31) for the random one. The symbols of corresponding colors show the numerical calculations performed for the SW model considering the same conditions. For both analytical solutions and numerical calculations, we assume $m_s = 340$ kA/m and $k_\mathrm{eff} = 3.2 \times 10^5$ J/m$^3$. The gray zone highlights the field range below $0.1 H_k$ in which we observe perfect convergence of the analytical solutions with the behavior predicted by the numerical calculations for the SW model.

regime ($H \ll H_k$), given by

$$m = m_s \cos\left(\alpha - \frac{\mu_0 m_s H}{2 k_\mathrm{eff}} \sin(\alpha)\right). \quad (29)$$

We may also access the solution for a random SW system averaging our findings over all possible orientations of $\vec{m}_s$,

$$m = m_s \left\langle \cos\left(\alpha - \frac{\mu_0 m_s H}{2 k_\mathrm{eff}} \sin(\alpha)\right)\right\rangle$$

$$= m_s \frac{\int_0^\pi \cos\left(\alpha - \frac{\mu_0 m_s H}{2 k_\mathrm{eff}} \sin(\alpha)\right) 2\pi \sin(\alpha)\, d\alpha}{\int_0^\pi 2\pi \sin(\alpha)\, d\alpha}, \quad (30)$$

which, for $H \ll H_c$, simplifies to

$$m \approx \frac{\mu_0 m_s^2}{3 k_\mathrm{eff}} H. \quad (31)$$

As a straightforward consequence, the initial magnetic susceptibility for a random SW system is

$$\chi = \frac{\mu_0 m_s^2}{3 k_\mathrm{eff}}. \quad (32)$$

To verify the validity of our findings, we first investigate the dependence of the magnetization $m$ on the magnetic field $H$ and compare the results obtained with the analytical solutions, given by Eqs. (29) and (31), with those achieved via numerical calculations performed for the SW model. Figure 2 shows the magnetization $m$ as a function of the magnetic field $H$, taking into account $\theta = 0$ as the initial situation, for selected $\alpha$ values. For both analytical solutions and numerical calculations, we consider $m_s = 340$ kA/m and $k_\mathrm{eff} = 3.2 \times 10^5$ J/m$^3$, which are typical parameters found for uniaxial anisotropy nanoparticles of barium hexaferrite, BaFe$_{12}$O$_{19}$ [45,49,50].

The magnetization curves present a clear dependence on the orientation between the easy magnetization axis and the magnetic field, reflecting all traditional features of uniaxial systems, as expected. From the plots, we clearly observe perfect convergence of the analytical solutions (solid lines) with the behavior predicted by the numerical calculations for the SW model (symbols) at the magnetic fields below $0.1 H_k$ (gray zone). The agreement between analytical solutions and calculations is understood to be evidence confirming that our approximation, i.e., Eqs. (29) and (31), is valid just at this low-field range.

Next, we explore the magnetic susceptibility from the behavior of the magnetization $m$ with the field $H$. However, given everything that has been stated above, from now on we focus our analysis on the magnetization regime at $H < 0.1 H_k$, thus informing the initial magnetic susceptibility.

Coming back to Eq. (27), it is worth observing that the maximum torque is found when $\vec{H}$ is perpendicular to the orientation of the uniaxial magnetic anisotropy axis of the particle. Hence, considering the solution for the magnetization provided by Eq. (29) and assuming $\alpha = \pi/2$, we obtain the maximum initial magnetic susceptibility $\chi_\mathrm{max}$ in the SW model, which is

$$\chi_\mathrm{max} = \left(\frac{\partial m}{\partial H}\right)_{H=0} = \frac{\mu_0 m_s^2}{2 k_\mathrm{eff}}. \quad (33)$$

It is interesting to note that Eq. (33) can be obtained directly from Eq. (8).

Moving forward, from Eqs. (25) and (33), we find the unexpected inequality

$$k_\mathrm{eff} \geqslant \tfrac{1}{2} \mu_0 m_s^2. \quad (34)$$

Equation (34) corresponds to the second fundamental inequality we aim to explore here. It is, in principle, valid only for noninteracting, uniaxial anisotropy single-domain (macrospin) blocked systems. This is a quite intriguing result indeed, especially if we realize that the $m_s$ and $k_\mathrm{eff}$ parameters in the context of the SW model are, a priori, independent. In other words, Eq. (34) correlates $k_\mathrm{eff}$ with $m_s$ and imposes a bottom limit for the magnetic anisotropy according to the saturation magnetization. Within this framework, one could read Eq. (34) as telling us that in systems described by the SW model, high saturation magnetization values in materials imply high uniaxial magnetic anisotropy, whatever the origin of such anisotropy is.

Figure 3 presents the dependence of the magnetic susceptibility $\chi_\mathrm{max}$ on the effective uniaxial magnetic anisotropy constant $k_\mathrm{eff}$, given by Eq. (33). For the analytical solution, we consider $m_s = 340$ kA/m. Notice that, once the saturation magnetization is set, there is a restricted region (gray zone) limiting the accessible values of $\chi_\mathrm{max}$ and $k_\mathrm{eff}$ for which the principle of energy conservation is not violated. For the initial magnetic susceptibility in a SW system, the upper limit is equal to 1, given by Eq. (25). Further, once the saturation magnetization is set here, Eq. (34) provides the bottom limit for the uniaxial magnetic anisotropy. For the case illustrated here, only $k_{eff}$ values larger than $0.73 \times 10^5$ J/m$^3$ are allowed.

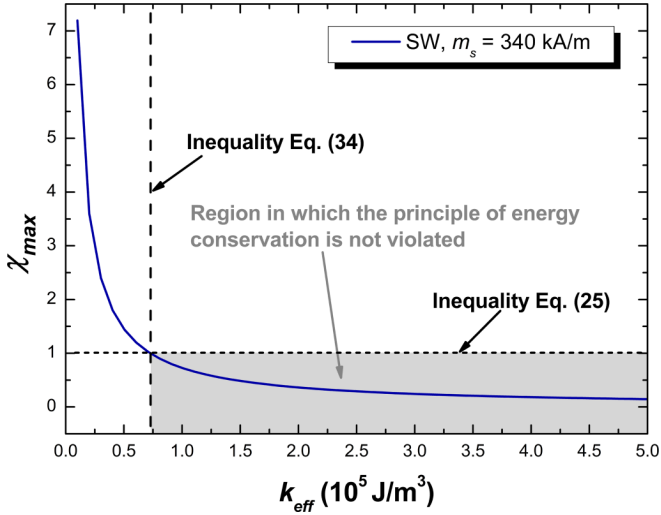

FIG. 3. Maximum magnetic susceptibility $\chi_{max}$ as a function of the effective uniaxial magnetic anisotropy constant $k_{eff}$. The $\chi_{max}$ values for the SW system are obtained using Eq. (33), assuming $m_s = 340$ kA/m. The horizontal dashed line represents the theoretical upper limit of the initial magnetic susceptibility in a SW system given by Eq. (25). The vertical dashed line corresponds to the bottom limit for the uniaxial magnetic anisotropy constant given by Eq. (34) once we set the saturation magnetization for the calculation. The gray zone highlights the magnetic susceptibility and uniaxial magnetic anisotropy constant ranges in which the principle of energy conservation is not violated.

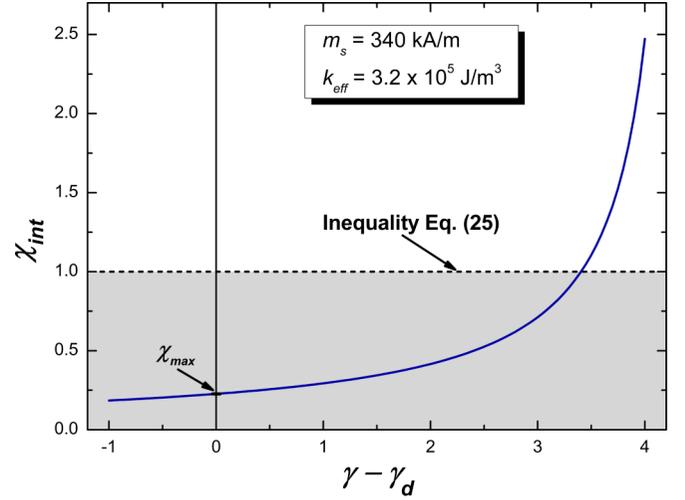

FIG. 4. Initial magnetic susceptibility $\chi_{int}$ for an interacting SW-like system as a function of the difference of the effective mean-field constant and effective demagnetizing factor $\gamma - \gamma_d$. The $\chi_{int}$ values for the interacting SW-like system are obtained using Eq. (37), assuming $m_s = 340$ kA/m and $k_{eff} = 3.2 \times 10^5$ J/m$^3$. Notice that only for positive $(\gamma - \gamma_d)$ values, in the vicinity of the divergence in the susceptibility, is the inequality in Eq. (25) violated.

### D. Violation of the first inequality in an interacting Stoner-Wohlfarth-like system

Going further, we determine whether the inequality in Eq. (25) is applicable for nonideal SW systems, i.e., interacting SW-like systems. To this end, we address here the case of a collective system of interacting magnetic nanoparticles. Let us assume that the sample is small enough, relative to the detecting system of the experimental setup, to be considered a point dipole. Notice that such a condition is often satisfied when the experiment is carried out by means of vibrating sample or superconducting quantum interference device magnetometers [45]. This assumption is a key factor that allows us to introduce in our theoretical approach the demagnetizing mean-field theory proposed by Sánchez and collaborators [33] and the Weiss's well-known mean-field theory [45,51,52]; as a consequence, both the dipolar and exchange interactions between magnetic nanoparticles inside the sample are inserted naturally in the model. From this perspective, the magnitude of the internal mean field $H_{in}$ in the sample may be expressed as

$$H_{in} = H + H_m + H_d = H + \gamma m - \gamma_d m, \quad (35)$$

where $H_m = \gamma m$ and $H_d = -\gamma_d m$ are the mean fields associated with the exchange and dipolar interactions, respectively, with $\gamma$ being the effective mean-field constant and $\gamma_d$ corresponding to the effective demagnetizing factor [33]. Here, we keep using the magnitudes $H$ and $m$, instead of the vectorial form.

Again, it is worth remembering that we understand that each nanoparticle behaves like a macrospin. This condition implies that the coherent magnetization rotation of each nanoparticle, i.e., the magnetic free energy associated with the exchange forces inside each single domain, is always minimized. Thus, taking into account the general mean-field theory represented by Eq. (35) and considering the situation in which the maximum magnetic susceptibility in the SW model is found, $\alpha = \pi/2$, the magnetization in the low-field regime ($H_{in} \ll H_k$) may be written from Eq. (29) as

$$m = m_s \sin\left(\frac{\mu_0 m_s H_{in}}{2k_{eff}}\right) \approx \frac{\mu_0 m_s^2}{2k_{eff} - \mu_0 m_s^2 (\gamma - \gamma_d)} H, \quad (36)$$

and consequently, the maximum magnetic susceptibility for an interacting SW-like system $\chi_{int}$ is

$$\chi_{int} = \frac{m}{H} = \frac{\mu_0 m_s^2}{2k_{eff} - \mu_0 m_s^2 (\gamma - \gamma_d)}. \quad (37)$$

Then, to evaluate the behavior of the magnetic susceptibility of an interacting SW-like system using the mean-field theory, Fig. 4 brings to light the dependence of $\chi_{int}$ on $(\gamma - \gamma_d)$. For this analytical solution, we consider $m_s = 340$ kA/m and $k_{eff} = 3.2 \times 10^5$ J/m$^3$. Notice the remarkable evolution of $\chi_{int}$ with $\gamma - \gamma_d$. The quantity $\gamma - \gamma_d$ discloses the type of interaction between the nanoparticles in the Stoner-Wohlfarth-like system. Specifically, negative $\gamma - \gamma_d$ values represent SW-like systems with interparticle interactions of dipolar origin; positive $\gamma - \gamma_d$ values indicate systems with ferromagnetic exchange interactions between nanoparticles, while $\gamma - \gamma_d = 0$ corresponds to the ideal noninteracting SW system.

The most striking finding here resides in the fact that for positive $(\gamma - \gamma_d)$ values, in the vicinity of the divergence in the susceptibility, $\chi_{\text{int}}$ achieves values above 1. In other words, the inequality in Eq. (25) is violated. Hence, although $\chi_{\text{int}}$ values smaller than 1 do not provide us insights allowing the identification of the existence and/or type of interparticle interaction, the violation of such an inequality is an unambiguous signature of the existence of ferromagnetic exchange interactions between nanoparticles in an interacting Stoner-Wohlfarth-like system.

## III. CONCLUSION

In conclusion, we derived two fundamental inequalities belonging to the Stoner-Wohlfarth model. First, we showed that the maximum work per unit volume that can be done by an electrical power source in the experimental setup is precisely the conservative energy per unit volume stored in the magnetic field in vacuum. Then, by means of the calculation of the work undergone by an ideal SW system at low fields, i.e., $H \ll H_c$ and $H \ll H_k$, we verified that the initial magnetic susceptibility may be written as the ratio of the work undergone by the system [Eq. (21)] to the maximum work that can be done by the electrical power source of the experimental setup in the magnetization process [Eq. (16)]. As result, we uncovered through a simple insight that $\chi \leqslant 1$ [Eq. (25)] for the magnetization curve in the linear regime at low fields. It is the first fundamental inequality we have addressed here, which is valid for any noninteracting single-domain (macrospin) blocked system and represents the theoretical limit of the initial magnetic susceptibility in an ideal nanoparticle system described by the SW model. Further, we also found analytical solutions for the magnetization in the low-field regime and obtained the borderline value for the uniaxial anisotropy constant in such an ideal Stoner-Wohlfarth system [Eq. (34)]. It corresponds to the second fundamental inequality we have disclosed here and correlates the $k_{\text{eff}}$ and $m_s$ parameters, in addition to imposing a bottom limit for the magnetic anisotropy according to the saturation magnetization (two parameters that are often understood as being independent). Going further, we finally introduced a general mean-field theory for interacting SW-like systems and estimated how the initial magnetic susceptibility is affected by the dipolar and exchange interactions between nanoparticles. Within this context, we demonstrated that the violation of the inequality $\chi \leqslant 1$ [Eq. (25)] is an unambiguous signature of the existence of ferromagnetic exchange interactions between nanoparticles in an interacting Stoner-Wohlfarth-like system.

TABLE I. Conventional SI units of the main quantities considered in the theoretical approach.

| Quantity | Unit |
| --- | --- |
| $\mu_0$ | $4\pi \times 10^{-7}$ N/A$^2$ |
| $E_Z, E_a, E_f$ | J |
| $m, m_s, m_r$ | A/m |
| $H, H_k, H_c, H_{\text{in}}, H_m, H_d$ | A/m |
| $k_{\text{eff}}, w, w_s, u_m$ | J/m$^3$ |
| $V_p, V_{\text{in}}$ | m$^3$ |
| $i$ | A |
| $N$ | dimensionless |
| $l$ | m |
| $A$ | m$^2$ |
| $B$ | T, N/(A m) |
| $\Phi$ | Wb, J/A |
| $t$ | s |
| $\varepsilon_b, V$ | V, W/A |
| $P$ | W |
| $X_L$ | $\Omega$, W/A$^2$ |
| $\omega$ | s$^{-1}$ |
| $L, L_0$ | H, J/A$^2$ |
| $\mu$ | N/A$^2$ |
| $\mu_r, \chi, \chi_{\text{max}}, \chi_{\text{int}}, \gamma, \gamma_d$ | dimensionless |

## ACKNOWLEDGMENT

This research is supported by the Brazilian agencies Conselho Nacional de Desenvolvimento Científico e Tecnológico (CNPq) and Coordenação de Aperfeiçoamento de Pessoal de Nível Superior (CAPES).

## APPENDIX

Table I gives the conventional SI units of the main quantities considered in this work.